\documentclass[lettersize,journal]{IEEEtran}
\usepackage{amsmath,amsfonts}
\usepackage{algorithmic}
\usepackage{algorithm}
\usepackage{array}
\usepackage[caption=false,font=normalsize,labelfont=sf,textfont=sf]{subfig}
\usepackage{textcomp}
\usepackage{stfloats}
\usepackage{url}
\usepackage{verbatim}
\usepackage{graphicx}
\usepackage{cite}
\begin{document}

\title{Attention-Enhanced Prioritized Proximal Policy Optimization for Adaptive Edge Caching}

\author{Farnaz Niknia, 
Ping Wang,~\IEEEmembership{Fellow,IEEE}
Zixu Wang,
Aakash Agarwal,
Adib S. Rezaei

\thanks{Farnaz Niknia and Ping Wang are with the Lassonde School of Engineering at York University, Toronto, ON, M3J 1P3, Canada (email: fniknia@yorku.ca, pingw@yorku.ca)}

\thanks{Zixu Wang is with the School of Electronic and Computer Engineering at the Hong Kong University of Science and Technology, Hong Kong, 999077, China (email: zwangjx@connect.ust.hk)}

\thanks{Aakash Agarwal is with the School of Engineering and Applied Science at the University of Pennsylvania, PA 19104, USA (email: aakash24@seas.upenn.edu)}

\thanks{Adib S. Rezaei is with the School of Electrical and Computer Engineering at the University of Tehran, Tehran, 1439957131, Iran (email: adib.rezaei@ut.ac.ir)}
}

\maketitle

\begin{abstract}
This paper tackles the growing issue of excessive data transmission in networks. With increasing traffic, backhaul links and core networks are under significant traffic, leading to the investigation of caching solutions at edge routers. Many existing studies utilize Markov Decision Processes (MDP) to tackle caching problems, often assuming decision points at fixed intervals; however, real-world environments are characterized by random request arrivals. Additionally, critical file attributes such as lifetime, size, and priority
significantly impact the effectiveness of caching policies, yet existing research fails to integrate all these attributes in policy design. In this work, we model the caching problem using a Semi-Markov Decision Process (SMDP) to better capture the continuous-time nature of real-world applications, enabling caching decisions to be triggered by random file requests. We then introduce a Proximal Policy Optimization (PPO)-based caching strategy that fully considers file attributes like lifetime, size, and priority. Simulations show that our method outperforms a recent Deep Reinforcement Learning-based technique. To further advance our research, we improved the convergence rate of PPO by prioritizing transitions within the replay buffer through an attention mechanism. This mechanism evaluates the similarity between the current state and all stored transitions, assigning higher priorities to transitions that exhibit greater similarity. 
 
\end{abstract}


\section{Introduction}

\IEEEPARstart{W}{ith} the rapid expansion of mobile applications, there has been a notable rise in redundant data transmissions due to numerous users requesting content from centralized data centers. This surge in traffic has significantly strained backhaul links and core networks \cite{DRL_reactive_caching}. Consequently, edge caching at routers has emerged as a promising strategy to alleviate traffic redundancy and reduce transmission delays, as demonstrated by recent research \cite{caching_info_centric_networking, dist_cache_2010}.

Current caching strategies can be classified into two main types: reactive and proactive \cite{app_cache_2021, video_aware_sched}. Reactive caching entails deciding whether to store a file only after it has been requested \cite{DRL_reactive_caching, ping_drl_iot, ahan_reactive_cache}. Proactive methods, in contrast, rely on historical data to forecast future file popularity and pre-cache the content even before any requests are made \cite{zhang2019proactivecache, jiang2019multiagent, rao2016optimal}. However, a key limitation of proactive caching is that it may result in low cache-hit ratios if the predicted content fails to become popular. Additionally, pre-caching content that may never be accessed wastes both communication bandwidth and valuable storage space, making reactive methods a more efficient alternative.

Another essential consideration is that caching inherently involves sequential decision-making, which is effectively modeled by the Markov Decision Process (MDP) framework, as outlined in \cite{puterman2014markov}. An MDP is defined by a set of states and actions available to an agent, who receives rewards based on their actions and the resulting states. The goal of the agent is to derive a policy that maximizes long-term expected rewards. If the model of the environment, such as the reward function and transition probabilities, is known, MDPs can be solved using Bellman optimality equations \cite{sutton2018reinforcement}. However, in many practical scenarios, such information is unavailable. In these cases, model-free Reinforcement Learning (RL) techniques are employed, allowing an agent to learn from the environment through exploration and trial-and-error rather than relying on pre-existing knowledge of the environment \cite{sutton2018reinforcement}.

The majority of existing studies apply the MDP framework to model caching problems \cite{wu2019dynamic}. However, MDP-based approaches assume decisions are made at fixed intervals \cite{sutton1999between}, which is not suitable for scenarios where caching decisions need to be made upon the arrival of requests. Since real-world environments feature requests arriving randomly at the edge router \cite{niknia2023edge}, MDP is often inadequate. A more fitting alternative is the Semi-Markov Decision Process (SMDP) \cite{puterman2014markov, sutton2018reinforcement}, which accommodates state transitions occurring at uneven intervals. Like MDP, SMDP provides a framework for modeling decision-making processes, but it offers more flexibility since state transitions can occur at equal, exponential, or random intervals \cite{puterman2014markov}. In fact, MDP can be seen as a special case of SMDP where the intervals between state transitions are one unit of time. The work proposed in \cite{zhang2022edge} exemplifies the application of SMDP for caching.

Another critical aspect of the caching problem is that not all files have the same level of importance. Importance here reflects how much a user values having a file stored nearby to avoid delays in future access. For instance, a user who frequently needs to access real-time financial data, such as live stock market updates, might be willing to pay a premium to ensure that this data is cached close by, allowing for immediate retrieval without latency. This approach provides a significant advantage to the caching system: by prioritizing files based on their importance to users, the system can enhance its overall benefit and efficiency. In addition to the importance of files, other attributes like size and lifetime also play a key role in optimizing caching policies. For instance, consider a scenario where a popular file is large but has a short lifetime. Caching this file might require evicting several other files from the cache. If the file expires before it is accessed again, this would not only make caching it ineffective but could also degrade overall system performance. Yet, most existing caching policies fail to account for all these attributes when making caching decisions.

Just as optimizing caching decisions requires careful consideration of these file characteristics, optimizing reinforcement learning algorithms also depends on making smart choices about which experiences to prioritize during training. Prioritizing transitions that closely resemble the current system state accelerates convergence by ensuring that the agent focuses on the most relevant experiences, directly informing the best policy updates. By concentrating on transitions closely related to its current state, the agent learns from experiences more applicable to its present situation, increasing the likelihood of policy effectiveness. This targeted approach reduces time spent on less relevant experiences, leading to more efficient learning and fewer updates needed for policy improvement. Moreover, learning from similar states minimizes variance in policy updates, stabilizing the learning process and resulting in smoother and more consistent progress toward convergence.

Motivated by these challenges, this paper proposes a reactive caching method built upon the SMDP framework \cite{puterman2014markov}, allowing for decision-making at random intervals, particularly when a file is requested at the edge router. Our main contributions are as follows:

\begin{itemize}
  \item{We model the caching problem using SMDP, which better reflects the real-time nature of request arrivals. We also present a PPO-based caching strategy that leverages historical popularity data to develop a caching policy while accounting for the system's inherent uncertainties.}
  
  \item{We incorporate multiple file attributes such as lifetime, size, and importance, along with popularity, in our caching decisions. To the best of our knowledge, our method is the first to integrate all these attributes into a comprehensive caching strategy, making it more applicable to practical environments.}

  \item{Through simulations, we assess our method's performance under various scenarios and compare it against two recent Deep Reinforcement Learning (DRL)-based approaches that consider both file popularity and lifetime \cite{zhu2018caching} and \cite{ping_drl_iot}. Results show that our approach consistently achieves a higher cache hit rate and total utility across different configurations, including varying cache sizes, request rates, and popularity distributions.}

  \item{We improve the convergence speed of the PPO algorithm by incorporating an attention mechanism to prioritize transitions in the replay buffer. This mechanism evaluates the similarity between the current state and all transitions in the replay buffer, assigning higher priority to those with greater similarity. This approach accelerates convergence by focusing learning on more relevant transitions.}
  
\end{itemize}

The remainder of the paper is structured as follows: Section \ref{Related Work} provides an overview of related work. Section \ref{System Model} introduces our system model, while Section \ref{Problem Formulation} formulates the caching problem. Section \ref{RL algorithm} outlines our proposed caching algorithm. In Section \ref{results}, we discuss our experimental setup and results, and Section \ref{Conclusion} concludes the paper.

\section{Related Work} \label{Related Work}

This section provides an overview of current caching techniques, covering both reactive and proactive strategies.

\subsection{Reactive Caching}

The study in \cite{ahan_reactive_cache} introduces a new metric called ‘virality', which measures the variation in file popularity over time. The authors use this metric along with popularity and size, to prioritize which files should be cached. Another approach detailed in \cite{DRL_reactive_caching} involves a recommendation system-based model to predict the popularity of newly requested content. This prediction guides a DRL-based caching strategy aimed at optimizing optimize caching decisions by balancing latency and request frequency.

In \cite{zhu2018caching}, the authors address the trade-off between communication costs and data freshness using an actor-critic DRL approach. They propose a utility function that combines these factors to improve caching efficiency. Similarly, in \cite{ping_drl_iot} the authors employ the PPO algorithm, aiming to enhance cache hit rates while minimizing energy use. A variant of this approach is presented in \cite{nasehzadeh2020deep}, where the DRL agent is penalized based on the age of cached files and available cache memory.

In \cite{yao2021cachedynamic}, the issue of average data transmission delay within cache storage constraints is tackled using deep reinforcement learning, initially formulated as an Integer Linear Programming (ILP) problem before applying DRL. The authors of \cite{zhong2020drlwireless} propose deep actor-critic methods for reactive caching, focusing on maximizing cache hit rates and managing transmission delays in both centralized and decentralized settings.

Study \cite{sun2023federateddeep} integrates recommender systems with edge caching in mobile edge-cloud networks, aiming to reduce the long-term system cost by modeling user experience factors. In \cite{xiaoge2023federatedlearning}, user preferences influence cache management, with higher preference content replacing lower preference items when space is limited. The study in \cite{gupta2023edgecomm} focuses on vehicular networks, using region-based models to optimize content fetching locations and employing the Least Recently Used (LRU) strategy for cache management.

\subsection{Proactive Caching}

Study \cite{wei2021wireless} addresses the 'slow start' problem in caching algorithms by calculating Euclidean distances to identify file similarities, assuming similar files are likely to be popular. In \cite{zhang2019proactivecache}, the caching of multi-view 3D videos is modeled models the caching of multi-view 3D videos using an MDP, combining Deep Deterministic Policy Gradient (DDPG) with a dynamic k-nearest neighbor (KNN) algorithm to adapt to varying action spaces.

In \cite{jiang2019multiagent}, the device-to-device (D2D) caching problem is formulated as a multi-agent multi-armed bandit problem, with Q-learning used to coordinate caching decisions among users. To manage extensive action spaces, a modified combinatorial Upper Confidence Bound (UCB) algorithm is employed. Study \cite{rao2016optimal} focuses on optimizing caching placement in a two-tier system, using Difference of Convex (DC) programming to maximize offloading probability through local caching and sharing.

Study \cite{zhou2023edgecomputation} proposes a computation offloading method that combines demand prediction using a Spatial-Temporal Graph Neural Network (STGNN) with a caching decision algorithm based on predicted demand. The authors in \cite{gao2021designdynamic} address varying content popularity by predicting average popularity and adjusting caching probabilities accordingly, managing different content popularities across runtime sessions.

Proactive caching faces challenges, such as potential low cache hit rates for pre-cached unpopular content and limited adaptability to dynamic user behavior changes. This can result in suboptimal performance if user preferences shift and previous caching decisions no longer apply.

\subsection{Motivation}

Real-world systems often contain files with diverse attributes that impact caching effectiveness, including lifetime, importance, and size. Our review reveals that current methods do not fully consider these attributes in caching decisions. Our proposed policy is the first to integrate a comprehensive set of file characteristics, offering a novel approach to caching.

Additionally, many existing methods model caching as a discrete-time problem \cite{zhang2019proactivecache} \cite{jiang2019multiagent} \cite{zhao2022intelligentcontent} \cite{DRL_reactive_caching} \cite{zhu2018caching} \cite{ping_drl_iot} \cite{wu2019dynamic} \cite{yao2021cachedynamic} \cite{zhong2020drlwireless}, which conflicts with the continuous-time nature of request arrivals. We address this by formulating the problem as an SMDP and developing a DRL-based reactive caching scheme. This approach maintains a brief request history and employs PPO to optimize the DRL agent's policy, enabling it to make adaptive caching decisions based on current and past experiences.

\section{System Model} \label{System Model}

\subsection{System Architecture}

\begin{figure}
  \begin{center}
  \includegraphics[width=2.5in]{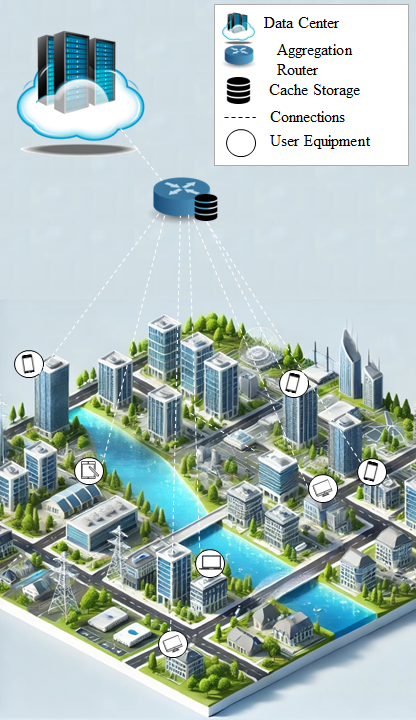}
  %\vspace{-15pt}
  \caption{Caching system topology}
  \label{fig System Model}
  \end{center}
\end{figure}

We consider a network scenario where end users access the Internet through an edge router, which connects to a data center on one side and the end users on the other side, as illustrated in Fig.~\ref{fig System Model}.

The cloud data center is assumed to have sufficient capacity to store all contents \cite{wei2021wireless}. When an end user requests a file, a copy is created and sent to the user via the edge router over the Internet. The edge router has a limited cache capacity of $M$ files. If the requested file $f_r$ is already cached, the request is fulfilled from the edge router's cache rather than retrieving it from the data center. Hereafter, we use “cache” to refer to the edge router's cache. End users are devices that request files based on their needs and preferences \cite{wei2021wireless}. Let $\mathcal{U} =\{u_1, u_2, \ldots, u_U\}$ denote the set of users connected to the edge router. Requests are represented by $\mathcal{G}=\{g_1, g_2, \ldots, g_G\}$, where $g_g$ denotes the $g^{th}$ request, irrespective of the user making it. Requests are processed in the order they are created.

% ==== FIG 1

Files, generated by sources such as cameras, sensors, and computers, are stored in the data center. Let $\mathcal{F}=\{f_1, f_2, \ldots, f_F\}$ denote the set of file types, where $f_f$ represents the $f^{th}$ type of file. Each file type may have distinct characteristics including popularity, lifetime, size, and importance. These characteristics are further explained with real-life examples:

\begin{itemize}
    \item \textbf{Popularity:} The number of times a file is requested by users \cite{li2016pop}. Popular files, such as trending videos, are requested more frequently than less popular ones.
    
    \item \textbf{Lifetime:} The duration for which a file is valid from the time it leaves the data center \cite{nasehzadeh2020deep}. For example, location-based services require timely updates, so files have a predetermined lifetime upon generation \cite{vural2017caching}.
    
    \item \textbf{Size:} Files may vary in size depending on their type and content. For instance, a movie file is typically larger than a text file.
    
    \item \textbf{Importance:} Reflects the value a file holds for users based on its relevance and necessity. Files that are crucial for timely access, such as real-time financial data or emergency updates, are considered more important than less critical files like free ebooks or casual videos.
    
\end{itemize}

A caching policy must consider all file characteristics simultaneously. Focusing on only one attribute may overlook other important features. For example, a highly popular file with a short lifetime, large size, and low importance might not be as beneficial as a less popular file with a longer lifetime, smaller size, and higher importance.

In this paper, we consider four characteristics for each file. Let $\mathbf{c}=(c_1, c_2, \ldots, c_F)$ denote the popularity values for each file type, where $c_f$ represents the popularity of file type $f$. This indicates the number of times file type $f$ is requested within a predefined time interval $\mathcal{T}$. We denote the lifetime, size, and importance of each file type at time $t$ by $\mathbf{l}=(l_1, l_2, \ldots, l_F)$, $\mathbf{z}=(z_1, z_2, \ldots, z_F)$, and $\mathbf{i}=(i_1, i_2, \ldots, i_F)$ respectively.

The utility of a cached file, defined as a function of its freshness and importance, is used to determine its value. Freshness is the age of the file normalized by its lifetime at time $t$:

\begin{equation*}
h^f(t) = \frac{t - w^f_g}{w^f_l}, \quad 0 \leq h^f(t) \leq 1,
\end{equation*}

\noindent where $t$ is the current time, $w^f_l$ is the lifetime, and $w^f_g$ is the generation time of file type $f$.

The utility function for file type $f$ at time $t$ is denoted as:

\begin{equation}
y_f(t) = E(h^f(t), i_f),
\label{utility function}
\end{equation}

\noindent where $y_f(t)$ represents the utility of file type $f$ at time $t$. The function $E$ depends on two key factors: the freshness of the file, denoted as $h^f(t)$, and the importance of the file type, represented by $i_f$. The function $E$ captures the interaction between these factors in determining the overall utility. In general, the utility $y_f(t)$ increases as the importance $i_f$ of the file grows, reflecting the higher benefit the file brings to the system. Notably, there is an inverse relationship between the file's freshness and its utility; as the freshness $h^f(t)$ decreases over time, indicating the file is becoming older, the utility $y_f(t)$ correspondingly. While the definition of a utility function often varies according to individual user perspectives and specific applications, there is no universally applicable definition. However, our proposed method is flexible and can be utilized with any definition of a utility function. 

\subsection{System Uncertainties}

Caching involves uncertainties such as random request arrivals and the unpredictable impact of future requests on the cache. These uncertainties impact the decision-making process of the edge router. Below, we outline these uncertainties and our assumptions.

\subsubsection{Random Arrivals of Requests}

Requests for content often arrive randomly due to factors like user behavior and network conditions. We model request arrivals using a Poisson process, a common model for characterizing user request patterns \cite{gomaa2013estimating}. The parameter $\lambda$ represents the request rate, and $\mu$ is the expected time between consecutive requests, where $\lambda = 1 / \mu$. We assume that the edge router has no prior knowledge of the Poisson process or its parameters.

\subsubsection{Unknown Effect of Upcoming Requests on the Cache}

Given the $F$ types of files with unique characteristics (popularity, lifetime, size, importance), there is no prior knowledge of future content requests. Consequently, caching a file may have an unforeseen impact on subsequent decisions.

These uncertainties significantly affect caching performance. In section \ref{RL algorithm}, we discuss how our caching policy addresses these uncertainties and propose an approach to optimize the hit rate in the long run.

\section{Problem Formulation} \label{Problem Formulation}

The edge router's task involves deciding which files to cache based on several factors, including the frequency of user requests, file characteristics, and the available cache space. This decision-making process is sequential and can be effectively modeled using MDPs, as detailed in \cite{puterman2014markov}.

An MDP consists of five components: state, action, system dynamics, reward function, and policy. In this framework, a state represents the system's status at a given time. The agent selects an action based on the policy, transitions to a new state, and receives a reward reflecting the action's quality. The process continues indefinitely (infinite horizon) or until a final state is reached (finite horizon) \cite{sutton2018reinforcement}. For the caching problem, the infinite-horizon case is appropriate since there is no final state.

Given that request arrivals are continuous, a continuous-time variant of MDP is necessary. We propose using the SMDP framework, which accommodates variable transition times \cite{puterman2014markov}. The following sections elaborate on the SMDP formalism and its components.

\subsection{SMDP Formalism}

An SMDP is defined by the tuple $(\mathcal{S}, \mathcal{A}, J, R, \pi)$, where $\mathcal{S}$ is the state space, $\mathcal{A}$ is the action space, $J$ represents transition times, $R$ is the reward function, and $\pi$ denotes the policy.

\subsubsection{States of the System}

The system state at time $t$ is denoted by $s_t \in \mathcal{S}$ and includes:

\begin{equation*}
s(t) = \{ Mem(t), \mathbf{b}(t), \mathbf{y}(t), \mathbf{d}(t), \mathbf{i}(t), \mathbf{l}(t), \mathbf{z}(t) \},
\end{equation*}

\noindent where $Mem(t)$ indicates the unoccupied percentage of cache memory, which is expressed as follows.

\begin{equation*}
Mem(t) = \frac{M-\sum_{f=1}^F{\mathbf{b}(t)\cdot \mathbf{z}(t)}}{M},
\end{equation*}

\noindent where $M$ is the cache capacity. The vector $\mathbf{b}(t) = (b_1(t), b_2(t), b_3(t), \ldots, b_f(t), \ldots, b_F(t))$, $\mathbf{b}(t) \in B = \{0, 1\}^F$, is a binary vector, where a value of 0 indicates that a file is not cached and a value of 1 indicates that the file is already cached. The vectors $\mathbf{y}(t) = (y_1(t), y_2(t), y_3(t), \ldots, y_f(t), \ldots, y_F(t))$, indicates the utility of files and $y_f(t) = 0$ if $f$ is not currently cached.  $\mathbf{d}(t)=\left(d_1(t), d_2(t), d_3(t), \ldots, d_f(t), \ldots, d_F(t)\right)$  represents the number of times that each file type has been requested within recent $N$ requests.  $\mathbf{i}(t)=\left(i_1(t), i_2(t), i_3(t), \ldots, i_f(t), \ldots, i_F(t)\right)$,  $\mathbf{l}(t)=\left(l_1(t), l_2(t), l_3(t), \ldots, l_f(t), \ldots, l_F(t)\right)$ and $\mathbf{z}(t) = (z_1(t), z_2(t), z_3(t), \ldots, z_f(t), \ldots, z_F(t))$ represent the importance, lifetime and size of each file type within recent $N$ requests. 

\subsubsection{Actions}

At any time $t$, the agent can take one of two actions: $a(t) = 1$ (cache the file) or $a(t) = 0$ (do not cache the file). When the cache is full, the file with the lowest utility is removed to make space for the new file. If more space is needed, files with the next lowest utilities are also removed.

\subsubsection{Dynamics of the System}

In our system, formulated as an SMDP with random task arrivals, the dynamics are governed by both the state transition probabilities $(P_{ss^\prime})$ and the transition times between states $(\tau)$. This makes the system's behavior more complex compared to standard MDPs, as both time and state transitions influence decision-making. If these dynamics, along with the reward function, were fully known, Bellman equations could be used to obtain the optimal policy. However, since such information is often unavailable in real-world problems, we employ reinforcement learning to iteratively learn these dynamics and optimize decision-making through experience. The specifics of the RL algorithm applied will be described in the next section.

\subsubsection{Instant Reward and Long-term Goal}

The instant reward is defined as:

\begin{equation}
r(t) = w_1 ((\mathbf{b}(t) \cdot \mathbf{d}(t))(\mathbf{b}(t) \cdot \mathbf{y}(t))^T) - w_2 Mem(t),
\label{reward function}
\end{equation}

\noindent where the first term is the weighted utility of cached files, and the second represents unused cache space. $w_1$ and $w_2$ are weighing coefficients.

The long-term goal is to maximize the average accumulated worth of cached files while minimizing the average unoccupied portion of the cache.

\subsubsection{The Policy}

The policy ($\pi$) determines the optimal action to take in each state to achieve the long-term goal.

\section{Enhanced PPO Algorithm with Prioritized Replay Buffer Using Attention Mechanisms} \label{Enhanced PPO with Prioritized Replay Buffer}

In this section, we introduce an enhanced version of the PPO \cite{schulman2017proximal} algorithm, which incorporates a prioritized replay buffer that leverages attention mechanisms. This enhancement aims to improve learning efficiency and policy performance by prioritizing transitions that are more pertinent to the current state of the agent. We first provide a detailed overview of the PPO algorithm, followed by an explanation of how the transitions are prioritized using attention mechanisms.

\subsection{Proximal Policy Optimization} \label{RL algorithm}

PPO is a popular reinforcement learning  algorithm known for its stability and ease of implementation. PPO addresses the challenges of policy optimization by introducing a clipped objective function, which prevents large policy updates and stabilizes training.

The core idea of PPO is to maximize the expected reward while ensuring that the new policy does not deviate excessively from the old policy. PPO achieves this through a surrogate objective function, which is given by:

\begin{equation}
L^{\text{PPO}}(\theta) =  \min \left( \frac{\pi_\theta(a(t) | s(t))}{\pi_{\theta_\text{old}}(a(t) | s(t))} \hat{A}(t), \right.
\end{equation}

\begin{equation*}
 \left. \text{clip} \left( \frac{\pi_\theta(a(t) | s(t))}{\pi_{\theta_\text{old}}(a(t) | s(t))}, 1 - \epsilon, 1 + \epsilon \right) \hat{A}(t) \right) ,
\end{equation*}

\noindent where $\pi_\theta(a(t) | s(t))$ is the probability of taking action $a_t$ in state $s_t$ under the policy parameterized by $\theta$, $\pi_{\theta_\text{old}}(a(t) | s(t))$ is the probability under the previous policy, $\hat{A}(t)$ is the advantage function, and $\epsilon$ is a clipping parameter.

PPO updates the policy by maximizing this objective function using stochastic gradient ascent. The clipped objective ensures that the new policy does not deviate significantly from the old policy, balancing exploration and exploitation.

The advantage function $\hat{A}(t)$ measures the relative value of an action compared to the baseline: 

\begin{equation}
\hat{A}(t) = r(t) + \gamma ^ \tau V_{\theta ^ \prime} (s^\prime) - V_{\theta ^ \prime}  (s),
\end{equation}

\noindent where $V_{\theta ^ \prime}  (s)$ and $ V_{\theta ^ \prime}  (s ^ \prime)$ represent the values of the current state and the next state, respectively. In the context of a standard MDP, $\tau$ is equal to 1 because state transitions occur at regular intervals, and the time between consecutive state transitions is fixed. As a result, $\gamma$ is raised to the power of 1, simplifying the equation. However, in SMDP, the state transition times vary, meaning that $\tau$ can take different values depending on the duration between transitions. In this case, the equation is modified by raising $\gamma$ to the power of $\tau$, which accounts for the variable time intervals between state transitions. This adjustment allows SMDP to more accurately reflect the delayed rewards over non-uniform time steps, ensuring that the future reward is appropriately discounted based on the actual time elapsed between state transitions \cite{SMDP}.

The value function is updated alongside the policy using a separate loss function. The loss for the value function is typically the squared error between the predicted value $V_{\theta ^ \prime}  (s)$ and the target return $R(t)$:

\begin{equation}
\label{original value loss}
L^{\text{value}}(\theta ^ \prime) =  \left( V_{\theta ^ \prime}  (s) - R(t) \right)^2 .
\end{equation}

This value loss ensures that the policy update is accompanied by an accurate estimation of state values. 

The target return \( R(t) \) is the one-step return defined as: 

\begin{equation}
R(t) = r(t) + \gamma^{\tau} V_{\theta ^ \prime}  (s ^ \prime).
\end{equation}

This formulation allows the value network to minimize the error between the current estimated value and the calculated return based on the immediate reward and future discounted value.

\subsection{Prioritizing Transitions with Attention Mechanisms}

To further enhance the PPO algorithm, we incorporate a prioritized replay buffer using an attention mechanism. 

\subsubsection{Attention Mechanism for Prioritization}

In our approach, we utilize the attention mechanism with the key-query-value (KQV) framework to compute the similarity between the current state and the transitions stored in the replay buffer. Let the current state be denoted as $s$, and the transitions in the replay buffer be represented as a set $(s_i, a_i, r_i, \tau_i, s_{i+1})$, where each transition in the replay buffer is characterized by its state $s_i$, action $a_i$, reward $r_i$, transition time $\tau_i$ and next state $s_{i+1}$.

We use the following steps to compute the attention-based priority for each transition:

\begin{itemize}
    \item \textbf{Compute Attention Scores}: Define the query $Q$ as the embedding of the current state $s$, and let the keys $K$ and values be the embeddings of the states in the replay buffer. The attention score $a_{i}$ for the transition $i$ is calculated using:

   \begin{equation}
   a_{i} = \frac{\exp(\text{score}(Q, K_i))}{\sum_{k=1}^{|\Lambda|} \exp(\text{score}(Q, K_k))},
   \end{equation}

   where $|\Lambda|$ is the size of the replay buffer. The score function $\text{score}(Q, K_i)$ measures the similarity between the query $Q$ and the key $K_i$. We use the dot product as the score function:

   \begin{equation}
   \text{score}(Q, K_i) = Q^T \cdot K_i.
   \end{equation}

    \item \textbf{Calculate Priorities}: The priority $p_i$ of a transition is proportional to its attention score. Define the priority as:

   \begin{equation}
   p_i = a_{i},
   \end{equation}

   \item \textbf{Update Probabilities}: The probability of transition $i$ being sampled is:

    \begin{equation*}
    P(i)=\frac{p_i^{\alpha}}{\sum_{e \in \Lambda} p_e^\alpha},
    \end{equation*}
    
    \noindent where the value of $\alpha$ determines the degree of prioritization, with $\alpha = 0$ corresponding to the uniform sampling case, i.e., transition $i$ is sampled randomly.
    To adapt to environment change, adjustments are made by applying importance sampling weights, as illustrated below:
    
    \begin{equation*}
    \omega_i=\left(\frac{1}{\left|\Lambda\right|} \cdot \frac{1}{P(i)}\right)^\beta,
    \end{equation*}
    
    \noindent where $\beta$ controls the degree of importance sampling. When $\beta$ is 0, there is no importance sampling, whereas when $\beta$ is 1, full importance sampling is employed.
    $\omega_i$ is multiplied by the loss function to control the impact of each transition on updating the neural network.

\end{itemize}

By integrating this attention-based prioritization into PPO, the algorithm evaluates the similarity between the current state and all stored transitions, assigning higher probability of sampling to transitions that exhibit greater similarity. By focusing on transitions with similar states with the current one, we ensure that the agent learns from experiences that are more relevant to its current situation. Prioritizing similar transitions leads to more efficient updates since transitions that resemble the current state are more likely to provide useful information for the decision-making process.

\subsection{The loss function}

We modify the loss function of the value network in Eq. (\ref{original value loss}) as follows:

\begin{equation}
L^{\text{value}}(\theta) = \mathcal{L}_{\text{1-step}} + \lambda_1 \mathcal{L}_{\text{n-step}} + \lambda_2 \mathcal{L}_{\text{reg}},
\end{equation}

\noindent
This updated loss function incorporates different components, including 1-step and n-step Temporal Difference (TD) losses, and an L2 norm regularization term. The parameters $\lambda_1$ and $\lambda_2$ control the relative contributions of each term to the overall loss function. By integrating these elements, the value network can effectively learn from both immediate and long-term rewards, while avoiding overfitting.

 The 1-step TD error, denoted as $\mathcal{L}_{\text{1-step}}$, is defined as:

\begin{equation*}
\mathcal{L}_{\text{1-step}} = \left( V(s) - R(t) \right)^2. 
\end{equation*}

The n-step TD loss, represented as $\mathcal{L}_{\text{n-step}}$, is expressed as:

\begin{multline}
\mathcal{L}_{\text{n-step}} = \left(R(t)_{(n)} + \gamma^{\tau_{(n)}} V_{\theta ^ \prime} (s^\prime_{(n)}) 
- V(s) \right) ^2,
\end{multline}

\noindent where $V_{\theta ^ \prime} (s^\prime_{(n)})$ refers to the state encountered after $n$ steps, and $\tau_{(n)}$ is the cumulative transition time from $s$ to $s^\prime_{(n)}$. The cumulative discounted reward $R(t)_{(n)}$ is calculated as:

\begin{equation*}    
\begin{gathered}
R(t)_{(n)}=\gamma^{\tau_{(1)}} r_{(1)}+\gamma^{\tau_{(2)}} r_{(2)}+\gamma^{\tau_{(3)}} r_{(3)}+\ldots+\gamma^{\tau_{(n)}} r_{(n)} \\
\tau_{(n)}=t_1+t_2+t_3+\ldots+t_n,
\end{gathered}
\end{equation*}

\noindent where $t_n$ represents the transition time from state $s^\prime_{(n - 1)}$ to $s^\prime_{(n)}$. $r_{(n)}$ is the immediate reward at the $n^{\text{th}}$ step. Incorporating n-step returns ensures that the values of subsequent states are propagated back to preceding states, enhancing the initial training process.

$\mathcal{L}_{\text{reg}}$ is an L2 regularization term designed to reduce overfitting by penalizing large weights. 

\section{Experimental Setup and Results} \label{results}
This section begins with a description of the experimental setup and the tools utilized to implement our Proposed Caching Algorithm (PCA). Next, we introduce two relevant and recent DRL algorithms, which will serve as baselines for comparison. 

Our system model was simulated and the DRL agent was trained using Python 3. To streamline the development of neural networks, we employed the TensorFlow platform as described in \cite{developers2022tensorflow}.

\subsection{Configuration and Parameters}
The DRL algorithm was implemented with 2 neural networks for actor and critic, each featuring 3 layers. The weights of the network were initialized within the range of [-0.1, 0.1], while biases were set to 0.1. ReLU served as the activation function. Additional parameter details can be found in Table \ref{table-results}.

\begin{table}[htbp]
  \centering
  \caption{parameter settings}
  \label{table-results}
  \begin{tabular}{lp{0.5\linewidth}}
    \hline
    \textbf{Notation}& \textbf{value}\\
    \hline
    $F$ & 50 \\
    $\gamma$ & 0.99 \\
    $M$ & 10000 \\
    $\beta$& 0.6 linearly increased to 1 \\
    $\alpha$ & 0.4 \\
    Batch size & 64 \\
    $|\Lambda|$ & 10000 \\
    $l$ for each file type & Randomly generated from [10, 30] \\
    $i$  for each file type & Randomly generated from [0.1, 0.9] \\
    $z$  for each file type & Randomly generated from [100, 1000] \\
    $\lambda$ & 0.2 \\
    $\eta$ & [0, 1] \\
    \hline
  \end{tabular}
\end{table}

File request probabilities follow a Zipf distribution, characterized by the parameter $\eta$ where $0 < \eta \leq 1$. In this distribution, the likelihood of the $f^{th}$ file being requested is given by $\mathfrak{p}_f = \frac{1}{\sigma f^\eta}$ \cite{gomaa2013estimating}, with $\sigma$ defined as:

\begin{equation}
\sigma = \sum_{f=1}^F {\frac{1}{f^\eta}},
\end{equation}

\noindent The value of $\eta$ influences the skewness of the Zipf distribution. When $\eta$ approaches 1, the likelihood of requesting the most popular file increases significantly compared to other files. Conversely, if $\eta$ is close to 0, the popularity of files becomes more evenly distributed. The utility function is set to increase linearly with the importance of a file and exponentially with its freshness.

\subsection{Baselines}

In this part, we introduce the baseline methods used for evaluating the performance of the proposed DRL approach.

\begin{itemize}
    \item \textbf{CTD}: Our DRL caching algorithm is compared against the approach presented in \cite{zhu2018caching}, where the authors modeled the caching problem as a discrete-time MDP and developed a DRL method to assist the edge router in deciding which files to cache based on various system states. In \cite{zhu2018caching}, file characteristics such as differing popularities and lifetimes were considered. However, other attributes like importance and file size were not included in their model. In our result comparisons, we refer to the approach in \cite{zhu2018caching} as CTD.

    \item \textbf{RLTD}: Another baseline we use for comparison is from \cite{ping_drl_iot}. This baseline proposes a DRL-based caching scheme, also considering the freshness and limited lifetime of data files. They proposed a distributed DRL approach for a hierarchical architecture where each cache in the hierarchy has a separate DRL agent operating independently. Since our work considers a single-level cache, we adapted this baseline to our system model by implementing it within a single-level cache. We present the results based on this configuration and call it RLTD.

\end{itemize}

\subsection{Evaluation Criteria}
Our proposed method is assessed using the following performance indicators:

\begin{itemize}
    \item \textbf{Cache Hits}: After training and policy convergence, different algorithms are tested with an additional 1000 user requests under identical conditions. Cache hits refer to the number of times a requested item is successfully retrieved from the cache. The primary objective of any caching strategy is to enhance this metric.
    
    \item \textbf{Total Utility}: When a requested file is found in the cache, its utility score is added to the cumulative utility. A higher aggregate utility signifies that the caching strategy effectively retains files with greater utility values.
    
    \item \textbf{Rate of Convergence}: An RL agent eventually stabilizes at its peak average reward. The efficiency of the algorithm is judged by how quickly it reaches this stable reward level. Rapid convergence is a key indicator of the learning algorithm's effectiveness.

\end{itemize}

\subsection{Experimental Findings}

This section begins with a comparative analysis of our simulation results against those of the CTD and RLTD methods, focusing on metrics such as total utility and cache hit count. We also compare the performance of our enhanced PPO algorithm with the original PPO algorithm to demonstrate how our enhancements lead to faster convergence. 

\paragraph{Effect of Popularity Skewness}

\begin{figure}
  \begin{center}
  \includegraphics[width=2.8in]{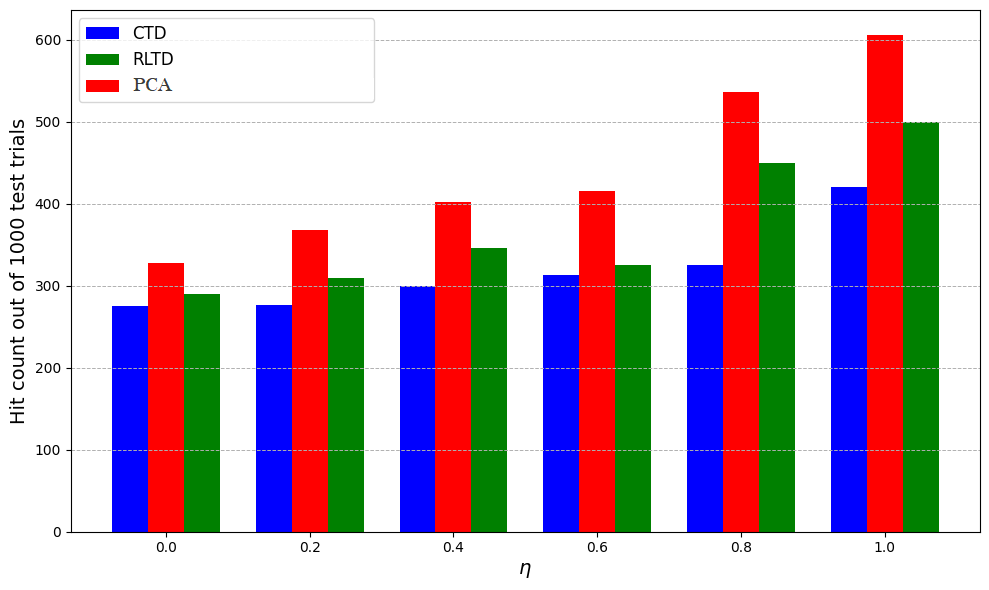}
  %\vspace{-15pt}
  \caption{Total hit counts for different values for $\eta$}
  \label{popularityhit}
  \end{center}
\end{figure}

\begin{figure}
  \begin{center}
  \includegraphics[width=2.8in]{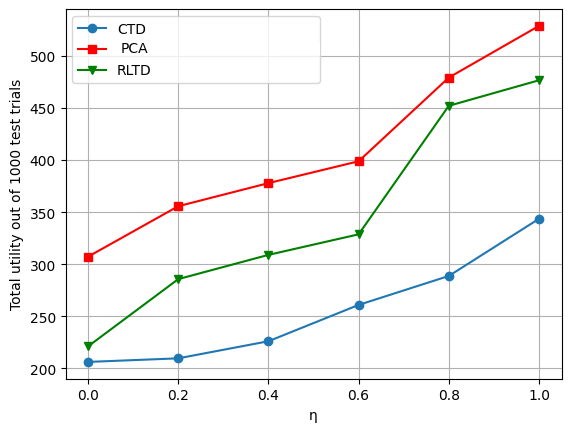}
  %\vspace{-15pt}
  \caption{Total utility for a) different values for $\eta$}
  \label{popularityUt}
  \end{center}
\end{figure}

Figures \ref{popularityhit} and \ref{popularityUt} illustrate the hit counts and total utility across 1000 trials for varying values of the Zipf parameter ($\eta$). The parameter $\eta$ governs the degree of skewness in the popularity distribution. A value of $\eta$ near 0 indicates that file popularities are relatively uniform, meaning there is minimal disparity between the most and least popular files. Conversely, when $\eta$ approaches 1, the distribution becomes highly skewed, with a few files dominating in popularity while others remain relatively obscure. This skewness means that caching the most popular files leads to a higher hit count, as increasing $\eta$ from 0 to 1 enhances the likelihood of such files being requested. Consequently, as $\eta$ rises, the total utility also increases due to the higher probability of retaining more frequently requested files, which are more likely to be up-to-date.

PCA outperforms both CTD and RLTD in terms of hit rate and total utility across all values of $\eta$, ranging from 0 to 1. This performance gap is due to PCA's ability to consider a wider range of file features, such as popularity, lifetime, importance, and size, which allows it to make more informed caching decisions. As the skewness $\eta$ increases, indicating a more concentrated demand for popular files, PCA optimizes cache usage by prioritizing high-utility files based on their combined features, not just popularity and lifetime like CTD and RLTD. 

\paragraph{Effect of Request Rates}

\begin{figure}
  \begin{center}
  \includegraphics[width=2.8in]{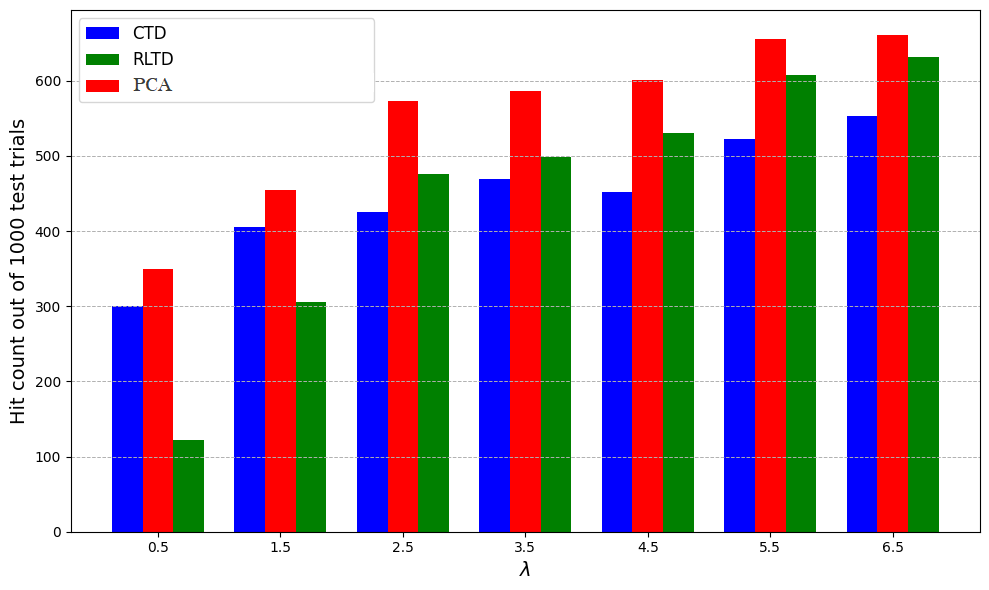}
  %\vspace{-15pt}
  \caption{Total hit count for different values for $\lambda$}
  \label{Interhit}
  \end{center}
\end{figure}

\begin{figure}
  \begin{center}
  \includegraphics[width=2.8in]{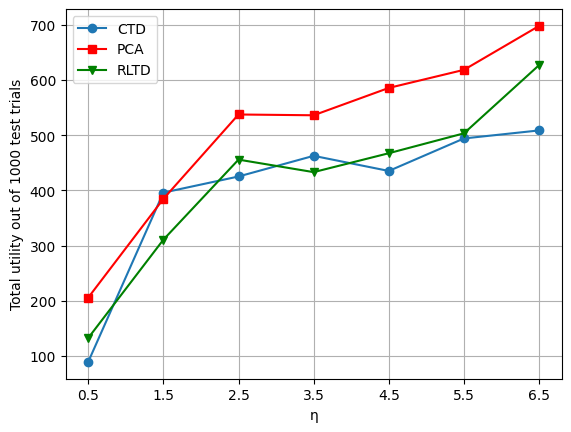}
  %\vspace{-15pt}
  \caption{Total utility for different values for $\lambda$}
  \label{InterUt}
  \end{center}
\end{figure}

Figures \ref{Interhit} and \ref{InterUt} depict the hit counts and total utility for varying request rates, denoted by $\lambda$, across 1000 trials. A lower $\lambda$ corresponds to longer intervals between consecutive requests, which is associated with reduced hit counts. This is because longer interarrival times increases the likelihood of cached files expiring before reuse due to their finite lifetimes. Conversely, a higher $\lambda$ means requests are more frequent, allowing the cache to fulfill more requests before files expire. This results in better performance in terms of both hit count and total utility. Our proposed method demonstrates superior performance compared to existing approaches for different values of $\lambda$.

\paragraph{Effect of Cache Size}

\begin{figure}
  \begin{center}
  \includegraphics[width=2.8in]{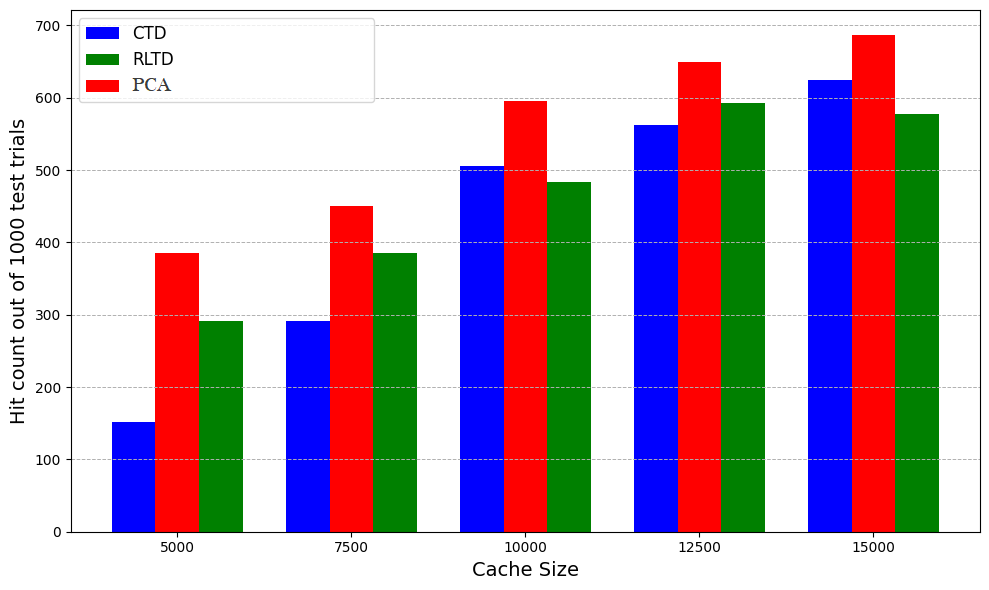}
  %\vspace{-15pt}
  \caption{Total hit count for different cache sizes}
  \label{CShit}
  \end{center}
\end{figure}

\begin{figure}
  \begin{center}
  \includegraphics[width=2.8in]{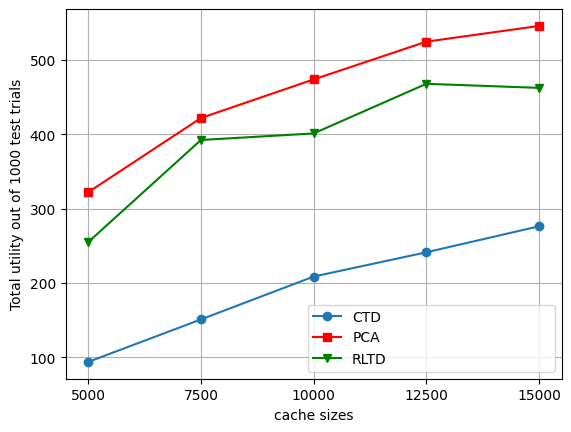}
  %\vspace{-15pt}
  \caption{Total utility for different cache sizes}
  \label{CSUt}
  \end{center}
\end{figure}

Increasing the cache size enhances the system's capacity to store more files, which generally leads to a higher number of requests being served directly from the cache rather than fetching them from the data center. This increased capacity results in improved hit counts with larger caches. As illustrated in Figure \ref{CShit}, our proposed approach consistently performs better than the benchmark caching strategies across various cache sizes. Additionally, as shown in Figure \ref{CSUt}, our method also surpasses the benchmark methods in terms of total utility.

\paragraph{Comparison of MDP and SMDP Approaches}

To showcase the benefits of using SMDP, we conducted an experiment with an alternative version of our caching algorithm, where the problem is framed using a discrete-time MDP instead of SMDP. In the discrete-time MDP setup, decisions are made at the beginning of each time interval, causing the agent to delay actions until the next time slot arrives, even if there are pending requests. Table \ref{SMDP and MDP comparison} shows the hit counts for the MDP and SMDP models. Since the agent in the MDP framework waits for a new time slot to make a decision, the files already in the cache continue to age, and they may even expire before being accessed. In contrast, SMDP allows the agent to make decisions as soon as a request arrives, optimizing the time that files remain cached for serving requests. As a result, the hit rate improves because cached files are utilized more efficiently. This increase in hit rate is especially significant when the request rate is higher, as more frequent requests with smaller interarrival times benefit from the SMDP's ability to respond immediately, reducing the likelihood of expired or outdated cached files.

\begin{table}[htbp]
  \centering
  \caption{Hit count for MDP and SMDP under different request rates}
  \label{SMDP and MDP comparison}  
    \begin{tabular}
    {|c|p{1.5cm} p{1.5cm} p{1.5cm}|}
    \hline
    & \textbf{$\lambda = 5$}& \textbf{$\lambda = 1.66$}& \textbf{$\lambda = 1.0$} \\      
    \hline
    \textbf{MDP} 
    & 405& 431& 399    \\
    \hline        
    \textbf{SMDP}& 610& 526& 443
    \\
    \hline
    \end{tabular}
\end{table}

Through extensive experimentation, we demonstrate that PCA significantly outperforms both CTD and RLTD in terms of cache hit rate and total utility under various scenarios. Additionally, we find that RLTD performs better than CTD. In the following sections, we explain the key reasons for these performance differences.

The superior performance of PCA can be attributed to two major contributions. First, PCA incorporates a comprehensive set of file features in its decision-making process, including file popularity, lifetime, importance, and size. In contrast, both CTD and RLTD limit their scope to only two features: file lifetime and popularity. By considering a richer set of features, PCA is better equipped to optimize caching decisions based on the specific characteristics of each file. This allows PCA to prioritize files not only based on their popularity but also on their importance and size, leading to a more efficient use of the cache and improved hit rates.

Second, PCA is formulated using SMDP, which enables decision-making at any moment when a request arrives at the edge router. This flexibility allows PCA to immediately respond to incoming requests and take action as needed, maximizing the use of cached files. On the other hand, both CTD and RLTD are based on a standard MDP, where decisions are only made at the beginning of fixed time intervals. As a result, when requests arrive between time intervals, the system must wait until the next interval to take action, which can cause delays in responding to requests and lead to lower hit rates. This limitation in MDP-based approaches explains why PCA, with its continuous decision-making capability, performs better, particularly in environments with high request rates and variable interarrival times.

The advantage of RLTD over CTD lies in their respective reward functions. CTD's reward function provides instant rewards solely based on the most recent requested file, which limits the feedback to the immediate outcome of the agent’s actions. In contrast, both RLTD and PCA use a more sophisticated reward function that takes into account the history of cached files. By considering a delayed reward system, these approaches better reflect the cumulative impact of previous caching decisions. This delayed feedback gives a more accurate evaluation of the agent’s long-term performance, rather than focusing only on short-term gains from responding to the most recent request. Consequently, RLTD and PCA are able to optimize caching policies more effectively over time, leading to better overall results compared to CTD.

\begin{figure}
  \begin{center}
  \includegraphics[width=2.8in]{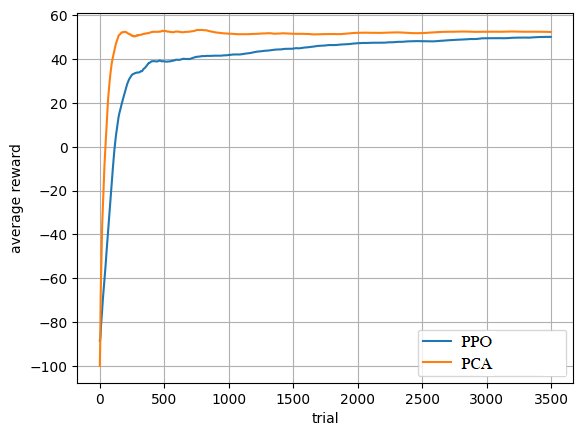}
  %\vspace{-15pt}
  \caption{PPO enhancement}
  \label{PPO Enhancement}
  \end{center}
\end{figure}

\paragraph{PPO Enhancement}
In our simulations, we observed a significant improvement in convergence speed with the enhanced PPO algorithm that incorporates experience prioritization using an attention mechanism in the replay buffer. The results are illustrated in Figure \ref{PPO Enhancement}, which compares the average reward over time during the training process for the standard PPO and PCA. The plot clearly shows that PCA converges much faster, achieving higher average rewards earlier in the training process compared to the baseline PPO. Particularly, the enhanced PPO algorithm reaches its stable average reward faster than the standard PPO.

The rapid convergence of the enhanced PPO can be attributed to the effective prioritization of experiences in the replay buffer. By leveraging the attention mechanism, the algorithm focuses on more relevant experiences, which accelerates the learning process and improves overall performance. This prioritization ensures that the agent learns from the most impactful transitions more frequently, facilitating a more efficient exploration of the state space. Consequently, this leads to quicker adaptation and refinement of the policy, thereby achieving superior performance in less time.

\section{Conclusion} \label{Conclusion}
In conclusion, our research presents a novel approach to caching challenges by leveraging an SMDP model to handle the real-world scenarios, where file requests arrive randomly at the edge router. Our proposed PPO-based caching method integrates a wide range of file attributes, such as popularity, lifetime, size, and importance. Simulation results highlight its superior performance compared to existing DRL-based methods, demonstrating improved efficiency. Additionally, we have enhanced the PPO algorithm by incorporating an attention mechanism to prioritize transitions in the replay buffer, leading to accelerated convergence and further improvements in performance.

\vfill

\end{document}